\documentclass[lettersize,journal]{IEEEtran}
\usepackage{amsmath,amsfonts}
\usepackage{amssymb}
\usepackage{textgreek}
\usepackage{algorithmic}
\usepackage{algorithm}
\usepackage{array}
\usepackage{xcolor}
\usepackage{booktabs}
\usepackage[caption=false,font=normalsize,labelfont=sf,textfont=sf]{subfig}
\usepackage{textcomp}
\usepackage{stfloats}
\usepackage{url}
\usepackage{verbatim}
\usepackage{graphicx}
\usepackage{lineno}
\usepackage{booktabs}
\usepackage{threeparttable}
\usepackage{tabularx} % For auto-wrapping columns
\usepackage[utf8]{inputenc}
\usepackage[style=ieee,doi=false,isbn=false,url=false]{biblatex}
\begin{document}

%\linenumbers

\title{Investigation of Using Non-Contact Electrodes for Fetal ECG Monitoring}

%Non-contact sensors-based device for antenatal continuous fetal monitoring}

\author{Tai Le,~\IEEEmembership{Student Member,~IEEE,} Hau Luu,~\IEEEmembership{Student Member,~IEEE,} Loan Pham-Nguyen,~\IEEEmembership{Member,~IEEE,} Hung Viet-Dao,~\IEEEmembership{Member,~IEEE,} Duc Nguyen Minh,~\IEEEmembership{Member,~IEEE,} Afshan B. Hameed, Hoang Nguyen, Liem Thanh Nguyen, Huy-Dung Han,~\IEEEmembership{Member,~IEEE,} Hung Cao,~\IEEEmembership{Senior Member,~IEEE}
        % <-this % stops a space
\thanks{This work was funded by Vingroup Joint Stock Company and supported by Vingroup Innovation Foundation (VINIF) under project code VINIF.2021.DA00138. The authors would like to thank members of BKIC Labratory at Hanoi University of Science and Technology for their support and insightful comments.

T. Le is with the Department of Biomedical Engineering, University of California Irvine, Irvine, CA, 92697, USA (e-mail: tail3@uci.edu).

H. Luu, L. P. Nguyen, H. V. Dao, D. N. Minh, H. D. Han is with the School of Electrical and Electronic Engineering, Hanoi University of Science and Technology, Hanoi, Vietnam \\(e-mail: hau.lt172537@sis.hust.edu.vn, loan.phamnguyenthanh@hust.edu.vn, \\hung.daoviet@hust.edu.vn, duc.nguyenminh5@hust.edu.vn, \\dung.hanhuy@hust.edu.vn).

Afshan B. Hameed is with the Department of Obstetrics \& Gynecology 
and Cardiology, University of California Irvine, CA 92697 USA (e-mail: 
ahameed@hs.uci.edu). 

Hoang Nguyen is with the Division of Pediatrics - Cardiology, UT Southwesten Medical Center, TX 75390 USA (e-mail: Hoang.Nguyen@UTSouthwestern.edu

Liem Thanh Nguyen is with Department of Research and Development, Vinmec Research Institute of Stem Cell and Gene Technology, Vinmec Healthcare System, Hanoi, Vietnam (e-mail: liem.nt@vinuni.edu.vn)

H. Cao is with the Departments of Electrical Engineering and Computer Science, Biomedical Engineering, and Computer Science, University of California Irvine, Irvine, CA, 92697, USA (e-mail: hungcao@uci.edu).
}}

\markboth{Journal of \LaTeX\ Class Files,~Vol.~14, No.~8, August~2021}%
{Shell \MakeLowercase{\textit{et al.}}: A Sample Article Using IEEEtran.cls for IEEE Journals}


\maketitle

\begin{abstract}
Regular physiological monitoring of maternal and fetal parameters is indispensable for ensuring safe outcomes during pregnancy and parturition. Fetal electrocardiogram (fECG) assessment is crucial to detect fetal distress and developmental anomalies. Given challenges of prenatal care due to the lack of medical professionals and the limit of accessibility, especially in remote and resource-poor areas, we develop a fECG monitoring system using novel non-contact electrodes (NCE) to record the fetal/maternal ECG (f/mECG) signals through clothes, thereby improving the comfort during measurement. The system is designed to be incorporated inside a maternity belt with data acquisition, data transmission module as well as novel NCEs. Thorough characterizations were carried out to evaluate the novel NCE against traditional wet electrodes (i.e., Ag/AgCl electrodes), showing comparable performance. A successful {preliminary pilot feasibility study} conducted with pregnant women (n = 10) between 25 and 32 weeks of gestation demonstrates the system's performance, usability and safety. 

\end{abstract}

\section{Introduction}
\IEEEPARstart{T}{h}ere are more than 200 million pregnant cases worldwide annually \cite{Bearak2020}. Although antenatal care has been improved to reduce neonatal and maternal morbidity and mortality, in the U.S., there are still 23.8 and 32.9 deaths per 100,000 live births in 2020 and 2021, respectively \cite{Hoyert03}. These numbers are significantly higher in low-resource countries with 157 deaths per 100,000 live births \cite{RN3}. Thus, assessing fetal development and well-being throughout pregnancy as well as having adequate prenatal care are essential in early detection of pregnancy anomalies, mitigating the risks of fetal/maternal mortality and/or morbidity with prompt and effective intervention. Traditionally, patients typically schedule appointments with healthcare professionals when symptoms manifest and adhere to their recommendations until the issue is resolved. This approach can be costly and time-consuming due to the need for frequent and prearranged visits to specialists.

Moreover, during the third trimester of pregnancy, cardiotocography (CTG) is required for monitoring fetal heart rate (fHR) and uterine contractions. CTG is the current standard of care for external monitoring of a fetus during a non-stress test (NST) and a contraction stress test (CST), as well as during labor. This method is intermittent and only provides a ‘snapshot’ of fetal condition and well-being, thus potentially failing to detect key signal of fetal demise \cite{RN4,RN5}. Additionally, CTG can be conducted only by a medical professional in the hospital setting as CTG Doppler sensors must be placed accurately for a robust signal and may need to be repositioned with fetal or maternal movement. Unsurprisingly, pregnancy monitoring with CTG is therefore largely limited in hospitals and clinics, which leads to the limit of accessibility. Given that fact, this standard of monitoring has changed minimally over the past 30 years; thereby indispensably calling for alternatives.

With recent advances in communications and technology, continuous fetal monitoring (CFM) offers the potential to alleviate these issues by providing an objective and longitudinal overview of fetal status and adequate care to expected mothers. Home-based pregnancy monitoring systems improve accessibility and thus help promote prenatal care. Most of these systems are based on recording abdominal ECG (the combined signal with both fECG and mECG) followed by signal processing and extraction to obtain individual fECG, mECG and their derivatives \cite{10474344, review_added_01, review_added_02}.
{Zhang et al. \cite{Zhang_et_al} introduced a portable ECG monitoring system. Their approach utilized Ag/AgCl adhesive electrodes for abdominal data collection and deployed an independent component analysis (ICA) algorithm to successfully isolate the fetal ECG component. Building upon this, the FetalCare system \cite{10472450} was proposed as a more integrated wearable solution, using adhesive stretchable electrodes to simultaneously capture both electrohysterography (EHG) and ECG signals for comprehensive fetal and maternal assessment. Despite these technological advances, a significant practical drawback remains. The FetalCare patch, for example, necessitates a complex arrangement of at least five electrodes on the abdomen, which can be inconvenient and limit user acceptance. Furthermore, the clinical validation for both systems was conducted on a very small scale—each with only three subjects—which limits the generalizability of their findings and raises questions about their performance on a wider population}. In parallel with these academic endeavors, the commercial market has also introduced several devices aimed at bringing fetal monitoring into the home. The Invu from Nuvo Inc. is designed as a wearable belt with multiple sensors to collect fHR and maternal HR as well as uterine contraction [4]. From Bloomlife Inc. (San Francisco, CA), Bloomlife Pregnancy Tracker is an innovative solution to capture EHG \cite{RN8}. Other systems, such as the Novii Wireless Patch System from  (Chicago, IL), and MERIDIAN M110 by MindChild Medical (North Andover, MA) are available to acquire fECG \cite{RN9}. These systems, however, are not sufficiently comprehensive to replicate full clinical monitoring as well as come with bulky packages (e.g., wire-connected, non-mobile). Some are limited to term pregnancies (at least 37 weeks of gestation) or the labor as well as delivery areas in the hospital.\\
Many pregnancy monitoring systems, both in commercial devices and academic research, have traditionally relied on conventional wet electrodes (i.e., Ag/AgCl electrodes). These are often used because they help obtain a stable signal with a high signal-to-noise ratio (SNR), which is crucial for reliable signal processing. However, this approach has significant drawbacks, particularly for long-term or continuous monitoring. The use of a conductive gel makes the electrodes inflexible for daily wear, and prolonged contact can cause skin irritation and allergic contact dermatitis, compromising user comfort and adherence \cite{RN10}. {This discomfort presents a significant barrier to the adoption of continuous, long-term monitoring in a home setting.} Recent technological advances enable dry and non-contact electrodes (NCE) based on capacitive measurement of electrophysiological measurement, which can acquire signals as high SNR as the conventional one and bring to users more convenient \cite{10908276,SOTA_01,SOTA_02}.\\
In this work, an antenatal fetal monitoring system using novel NCE sensors is first introduced to collect fetal/maternal ECG. {The novelty of this work lies in the synergistic integration of (a) an NCE with an improved, cost-effective bootstrap biasing circuit for the high-impedance requirements of fECG and input capacitance cancellation technique to ensure stable system gain, (b) its seamless incorporation into a comfortable maternity belt for practical long-term use, and (c) a pilot clinical feasibility study to validate its performance against standard Ag/AgCl electrodes}. This integrated monitoring system leverages advanced sensing technology, wireless connectivity, and compatibility with a wide range of low-cost mobile devices. The system includes a wearable belt, NCE sensors and an Android mobile application to receive data using Bluetooth Low Energy (BLE) communication. Comprehensive experiments are conducted to characterize the NCE sensors. A successful field trial in 10 pregnant subjects at UCI medical center demonstrates the system’s performance and usability.

\section{NCE Fetal ECG Monitoring Device}
The proposed NCE-based antenatal fetal monitoring system is designed to be integrated inside a maternity belt which consists of NCE sensors, a data acquisition module, and a signal transmission module in connection with a mobile phone app. The following sections  describe more detail for each component of the system.
\begin{figure}[!ht]
    \centering
    \includegraphics[width=\linewidth]{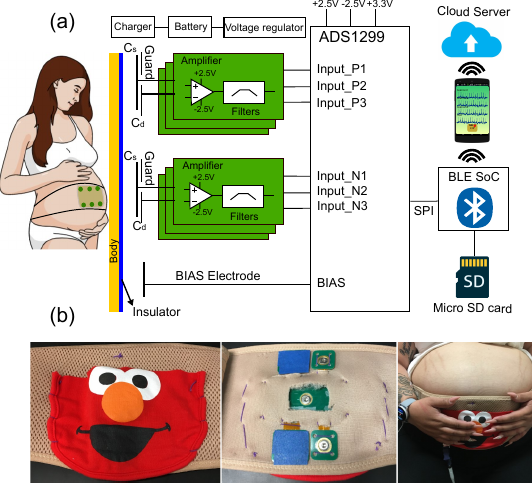}
    \caption{NCE-based antenatal fetal monitoring system: (a) System overview with data acquisition and mobile application; (b) Maternity belt with the system integrated: front side, back side and measurement setup on a pregnant women to collect data (from left to right).}
    \label{fig: Overview system}
\end{figure}

\subsection{Data acquisition and mobile application}
The center of the data acquisition is an analog front end of ADS1299 (Texas Instruments) with 24-bit analog-to-digital converter specifically designed for biomedical signal measurement. A system-on-chip nRF52832 (Nordic Semiconductor, Trondheim, Norway) powered with Arm Cortex-M4 CPU running at 64 MHz is utilized to transmit data from ADS1299 to the Android application through BLE as shown in Fig. \ref{fig: Overview system}a. 

Serving as a user interface, an Android smartphone application in Java was developed to connect to the Bluetooth Low Energy (BLE) device for data collection, displaying, and saving. Through BLE protocols, the application connects to the system and reads multiple channels of data at a rate of 500 Hz. After accumulating 1000 data points, the input data is sent to the connecting cloud server for feature extraction of fECG. The results will start appearing on the application interface after at least 3 seconds of initial start of data acquisition in the form of dynamic graphs. Moreover, a micro SD card was also used to store the data collected from the system, ensuring data integrity even if BLE connection has some issues.

Fig. \ref{fig: Overview system}b illustrates the whole system integrated in a maternity belt. A fabric with a smiling face is used to cover whole electronic components and battery while four slits made in the back of the belt is for four electrodes to contact the abdominal area. The complete belt then is worn on a pregnant subject.

\subsection{Design of non-contact electrode (NCE)}

\begin{figure}[!ht]
    \centering
    \includegraphics[width=\linewidth]{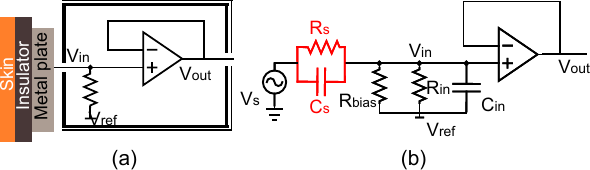}
    \caption{a) General scheme of a non-contact electrode b) non-contact electrode's equivalent electrical circuit}
    \label{fig: overall NCE}
\end{figure}

The basic concept of the proposed novel non-contact electrode is a conducting plate covered by an insulating layer to form a parallel plate capacitor when attached to the skin. The NCE can couple bio-signals capacitively to an amplifier. However, with the coupling capacitance between the skin and the conducting plate are of the order of pF, the impedance of skin-electrode interface is very high. It will cause the non-contact electrode to become very sensitive to interference; thereby, calling for ultra-high input impedance amplifiers and careful biasing, guarding and shielding techniques \cite{RN11}. Fig. \ref{fig: overall NCE}a illustrates the basic scheme of the proposed non-contact electrode with the main component of an amplifier circuit and a metal plate. A high-value resistor is connected with a reference voltage $V_{ref}$ to provide a bias voltage to ensure the operational amplifier works in the active region. 

The equivalent model of the proposed non-contact electrode is then shown in Fig. \ref{fig: overall NCE}b. Here, the interface of the sheet metal and the skin can be considered as a pair of capacitors $C_{s}$ and resistors $R_{s}$ in parallel. The values of these components vary with each electrode type, with non-contact electrodes, the influence of $C_{s}$ is predominant. Here, the bio-signal source is represented as $V_{s}$, $R_{bias}$ represents the equivalent impedance of the bias line, $R_{in}$ and $C_{in}$ are the amplifier input impedance and capacitance, respectively. The electrical characteristics of the model affect the signal quality at the input of the amplifier. The transfer function defining the relationship between the signal at the skin surface ($V_{s}$) and the signal at the input of the amplifier ($V_{in}$) is given as equation (\ref{equation - 1}):

\begin{equation}\label{equation - 1}
     V_{in}\simeq \frac{R_{in}//R_{bias}}{R_{in}//R_{bias}+R_{s}}.\frac{C_{s}}{{C_{s}+C_{in}}}V_{s}.
\end{equation}
\begin{figure}[!ht]
    \centering
    \includegraphics[width=\linewidth]{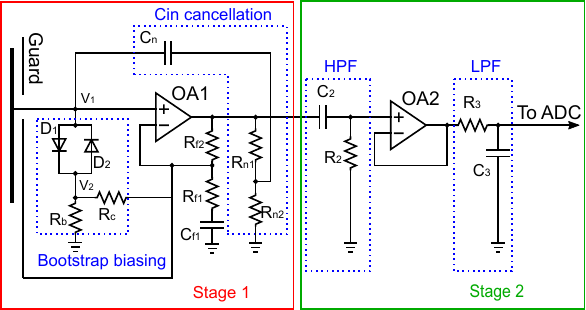}
    \caption{Full schematic of non-contact electrode with bootstrap biasing and $C_{in}$ cancellation technique.}
    \label{fig: NCE schematic}
\end{figure}
Fig. \ref{fig: NCE schematic} describes the novel non-contact sensor developed in this work, including two amplification stages. The first stage is to drive signal from body with minimum amplitude reduction throughout bootstrap biasing and $C_{in}$ cancellation technique. The second stage is to bring the signal through two analog filters, including a high pass filter (HPF) with the cutoff frequency of 0.07 Hz and a low pass filter (LPF) with the cutoff frequency of 250 Hz before feeding it to the ADC chip for digitization.

For the first stage, when using insulating electrodes, a path must be provided for the bias current of the operating amplifier (OA) as it cannot flow through the patient, and having this current path makes the electrode operate stably \cite{RN11,RN12}. However, the implementation of bias path should not reduce the input impedance of the OA or generate excessive noise. As an improvement from our previous works \cite{IEEEsensor,IMBioC, SOTA_02}, a pair of anti-parallel diodes was used, thus providing ultra-low leakage currents which is below 1 pA, or equivalently, providing the resistance larger than 50 G$\Omega$. This approach offers a cost-effective solution, as resistors exceeding 50 G$\Omega$ are quite expensive (over $\$100$). Furthermore, these resistors exhibit broad tolerance ranges and are vulnerable to variations in temperature and humidity \cite{Flavio}, so using a resistor as a bias path will degrade the performance of first stage. As shown in Fig. \ref{fig: NCE schematic}, two diodes $D_{1}$ and $D_{2}$ were wired in parallel and bootstrapped with $R_{b}$ and $R_{c}$. These diodes ensure input DC bias goes through GND and the bootstrap keeps dynamic voltage on diodes are nearly zero. We have an equivalent resistance $R_{d}$ of 2 diodes and the bias circuit equivalent resistor ($R_{bias}$) as shown in equation (\ref{equation - 2}) and (\ref{equation - 3}):
 
 \begin{equation}\label{equation - 2}
   R_{d}=\frac{V_{1}-V_{2}}{I_{D1}+I_{D2}},
\end{equation}

\begin{equation}\label{equation - 3}
   R_{bias}=R_{b}+R_{d}+R_{d}\frac{R_{b}}{R_{c}}\simeq R_{d}\frac{R_{b}}{R_{c}}.
\end{equation}
where, $I_{D1}$ and $I_{D2}$ are the current flowing through the diodes $D1$ and $D2$, respectively. By generating a very high biasing resistance ($R_{bias}$), the bootstrap circuit ensures that the input coupling capacitor ($C_{s}$) does not attenuate the crucial low-frequency components of the ECG signal, thus maintaining its full bandwidth.

Another factor that should be taken into account is the amplifier’s input capacitance as it couples with $C_{s}$ to establish a capacitive voltage divider. This causes the input signal’s attenuation, which will be addressed in this section. The first stage was set with a gain of 11 calculated by the equation (\ref{equation - 4}) with $R_{f1}$ = 1 $M\Omega$, $R_{f2}$ = 100 $k\Omega$  and $C_{f1}$ = 100 $\mu$f. With the high value of $C_{f1}$, it prevents amplification of low frequency components below the signal band. %With the high value of $C_{f1}$, it avoids the DC current, which allows it to remove input capacitance in whole frequency range of ECG signal.
\begin{equation}\label{equation - 4}
    A_v=1+\frac{R_{f1}}{R_{f2}+Z_{Cf1}}\simeq 1+\frac{R_{f1}}{R_{f2}},
\end{equation}
where $Z_{Cf1}$ is the corresponding impedance of $C_{f1}$. 
Here, the input capacitance neutralization circuit uses a voltage divider ($R_{n1}$ and $R_{n2}$) to adjust the positive feedback with the gain of $\alpha = \frac{R_{n2}}{R_{n1}+R_{n2}}$ which was connected to the op-amp’s input via neutralization capacitor $C_{n}$. The Miller effect allows for a reduction of the input capacitance $C_{in}$ to $C^{'}_{in}$, as illustrated in equation (\ref{equation - 5}) \cite{RN11}, thereby stabilizing the circuit gain without being influenced by $C_{s}$. 
\begin{equation}\label{equation - 5}
    C^{'}_{in}=C_{in}-\left(\frac{A_{v}R_{n2}}{R_{n1}+R_{n2}}-1\right)C_n.
\end{equation}

{The values for the voltage divider resistors ($R_{n1}$, $R_{n2}$) were determined by first calculating the theoretical ratio required using equation (5) and the performing empirical fine-tuning on the benchtop to achieve the most stable gain across the tested range of source capacitances (5-100 pF).} By carefully adjusting $R_{n1}$ and $R_{n2}$, it is possible to arrive at a value of $\alpha$ such that the input capacitance is almost completely negated, resulting in an amplifier gain that is essentially invariant with respect to $C_{s}$. The Laplacian transfer function $H(s)$ for our novel NCE with bootstrap biasing and $C_{in}$ cancellation is shown in equation (\ref{equation - 6}):
%\begin{equation}\label{equation - 6}
%    \footnotesize H(s) = {{\frac{sC_s[sC_{f1}(R_{f1}+R_{f2})+1]}{sC_s(sC_{f1}R_{f2}+1)+\frac{sC_{f1}R_{f2}+1}{R_{bias}}}}. \frac{s}{s+\frac{1}{R_2C_2}}.{\frac{1}{1+sR_3C_3}}}
%\end{equation}

\begin{eqnarray}\label{equation - 6}
    H(s) = {\frac{sC_s[sC_{f1}(R_{f1}+R_{f2})+1]}{s(C_s+C^{'}_{in})(sC_{f1}R_{f2}+1)+\frac{sC_{f1}R_{f2}+1}{R_{bias}}}} \nonumber \\
    \cdot \frac{s}{s+\frac{1}{R_2C_2}} \cdot {\frac{1}{1+sR_3C_3}}
\end{eqnarray}

With very small $C^{'}_{in}$ and very large $R_{bias}$ values achieved, NCE has approximately 11 times gain set by the first stage and bandwidth depend mainly on cutoff frequency of the second stage filters. This indicates that the proposed design has efficient amplification in the ECG signal band.

\section{Experimental Methods}
\subsection{Non-contact electrodes characterization}
During benchtop testing, the sensor underwent characterization through a series of experiments. Initially, the $C_{in}$ cancellation circuit was activated and deactivated to assess the circuit's gain consistency. For examining the skin and electrode surface interface behavior, multiple values of the source capacitance were applied. Subsequently, the gain and phase response were examined by following the process as depicted in Fig. \ref{fig: Cin cancellation}a. A sine-wave signal with a 300 mV peak-to-peak amplitude served as the input for the non-contact electrode. The sine wave's frequency ranged from 0.1 Hz to 10 kHz, generated by a function generator. The function generator output was connected to a metal plate, linking the conducting plate of the non-contact electrode through various source capacitance. The data acquisition card recorded the output signal of the non-contact electrode, which was then transmitted to a computer for the calculation of electrical characteristics. Finally, noise analysis was conducted by positioning two sensors facing each other with different source capacitors connected. The output data were collected, and the noise spectra were computed using MATLAB's $pwelch$ function to estimate the power spectral density (PSD) of the data sample.

\subsection{Experiment Protocol}
The electrocardiogram signal was captured from four healthy subjects using a system equipped with non-contact electrodes. In the initial phase, various insulators were employed, such as masking tape with different layers and cotton through a T-shirt. Each subject sat comfortably on a chair and underwent a two-minute ECG measurement. Throughout the experiment, participants were instructed to randomly hold their breath for brief periods. The subsequent experiment involved the same subjects assuming different postures (lying supine, sitting, and slow walking), with ECG measurements performed for each posture. 

Finally, a pilot study, approved by the Institutional Review Board (IRB) at the University of California, Irvine under IRB ID HS\# 2020-6342 (updated on 02/07/2023), was conducted on ten pregnant subjects. Informed consent was obtained from all participants involved in the study. The objective of this preliminary study was to confirm the capability of the innovative sensors in detecting a composite ECG signal generated by both the mother's and fetus's ECG. Consequently, each pregnant participant was requested to sit comfortably in the medical examination chair. As illustrated in Fig. \ref{fig: Overview system}b, this device was integrated into a maternity belt with  two sets of electrodes, comprising a pair of Ag/AgCl electrodes and a pair of non-contact electrodes that contact with pregnant subject's abdomen. In order to enhance the chances of acquiring data, we repositioned the electrodes across various locations on the abdominal region, encompassing the central, left, and right positions relative to the location of the belly button. {This practical, ad-hoc placement strategy was employed to maximize the likelihood of detecting the fECG signal without the use of ultrasound guidance}.

\subsection{Signal processing and Statistics}
Data obtained from the mobile app was transferred to a laptop for signal processing using MATLAB. The signal underwent high-pass and low-pass filtering with cutoff frequencies set at 3 Hz and 250 Hz, respectively. {This passband was selected based on established fECG processing standards; the 3 Hz high-pass filter effectively removes low-frequency baseline wander from sources like maternal respiration, while the 250 Hz low-pass filter eliminates high-frequency noise (e.g., muscle artifacts) outside the diagnostic band of the fECG signal}. Subsequently, Wavelet transform was applied to the data before proceeding to feature detection \cite{Le}.

Statistical analysis was conducted using OriginLab 2019 and involved several tests. Multiple comparisons were assessed through one-way ANOVA, and statistically significant results ($P < 0.05$) underwent further analysis using pairwise comparisons with Student's t-test, applying significance levels adjusted using the Bonferroni method. Significance was denoted with asterisks (*), with $^{*}P < 0.05$. Correlation analysis was carried out using Pearson's correlation.

In the pilot study, to evaluate the performance of the use of NCE sensor over the Ag/AgCl electrodes, the accuracy of fetal QRS (fQRS) detection in the data recorded by two different electrodes was assessed. {The simultaneously recorded Ag/AgCl signal served as the ground-truth reference, with fQRS locations being annotated by an expert}. A 100 ms window (±50 ms) is centered on each expert-annotated fQRS location. A detected fQRS is considered correct if it falls within this window \cite{Behar_2016} 
%\cite{Porr2024}
. The accuracy is then evaluated by using sensitivity (Se), positive predictive accuracy (PPV), accuracy (ACC) and their harmonic mean (F1) as defined below:
\begin{equation}
Se = \frac{TP}{TP+FN},
\end{equation}

\begin{equation}
PPV = \frac{TP}{TP+FP},
\end{equation}

\begin{equation}
Acc = \frac{TP}{TP+FN+FP+TN},
\end{equation}

\begin{equation}
F_1 =2 \cdot \frac{Se \cdot PPV}{Se + PPV} = \frac{2 \cdot TP}{2 \cdot TP + FN + FP},
\end{equation}

\noindent where $TP$ (True Positives), $FP$ (False Positives), and $FN$ (False Negatives) represent the quantities of correctly detected fQRS complex waves, incorrectly detected fQRS complex waves, and missed detections of fQRS complex waves, respectively.
\begin{figure}[!ht]
    \centering
    \includegraphics[width=\linewidth]{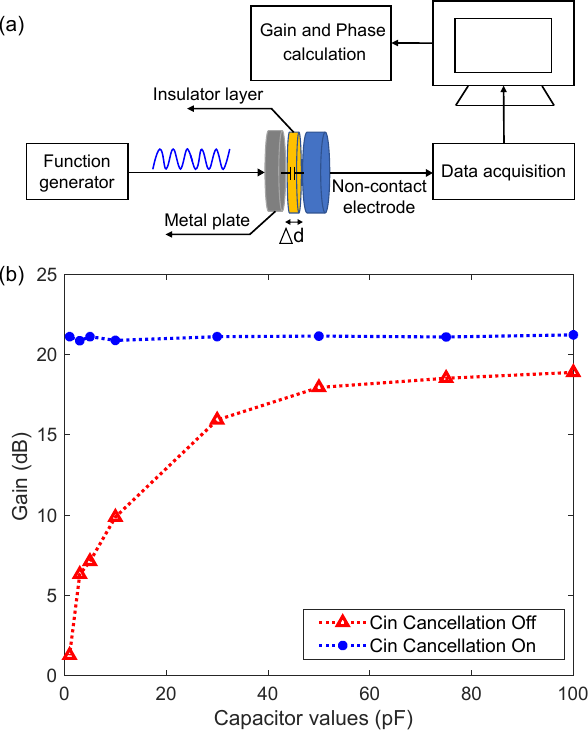}
    \caption{(a) Experiment illustration for phase and gain response calculation; (b) gain versus different source capacitance with and without $C_{in}$ cancellation technique.}
    \label{fig: Cin cancellation}
\end{figure}

\section{Results and Discussion}
\subsection{Non-contact electrode}

The NCE gain versus various source capacitance was measured before and after applying the $C_{in}$ cancellation technique as shown in Fig. \ref{fig: Cin cancellation}b. It is evident that the amplifier's full gain cannot be achieved without the $C_{in}$ cancellation circuit, even with the highest source capacitance of 100 $\mu$F which is equivalent to maximum coupling strength when sensor has a very thin insulation layer. When the $C_{in}$ cancellation technique was applied, the sensor gain remained stable across different source capacitance. This indicates that the sensor can be placed further from the patient without significant attenuation of the signal.

\begin{figure}[!ht]
    \centering
    \includegraphics[width=\linewidth]{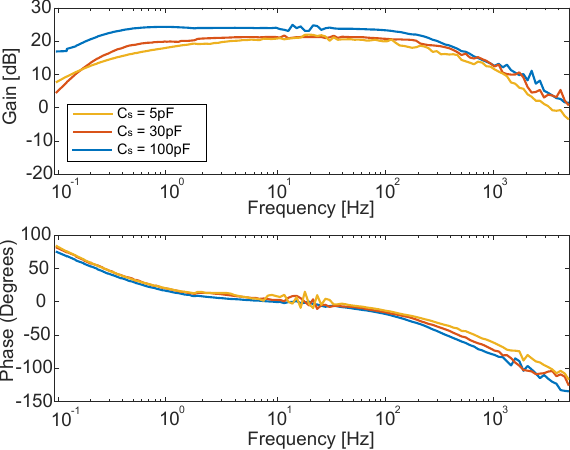}
    \caption{Phase and gain response with different source capacitors.}
    \label{fig: Bode plot}
\end{figure}

\begin{figure}[!ht]
    \centering
    \includegraphics[width=\linewidth]{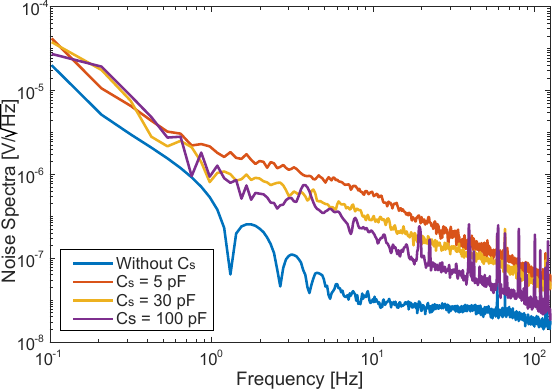}
    \caption{Effect of sensor separation distance being mimicked by different capacitor values on input-referred noise.}
    \label{fig: Noise analysis}
\end{figure}

Fig. \ref{fig: Bode plot} illustrates the frequency response of the sensor spanning from 1 Hz to 10 kHz for varying source capacitance ($C_{s}$) of 5 pF, 30 pF and 100 pF. The response achieved with the maximum source capacitance of 100 pF maintained a consistently flat frequency response, closely resembling the ideal gain obtained at the amplifier. Although lower source capacitance exhibited a slight decrease in gain across the frequency range, their phase responses remained consistent. This implies that the sensor input boasts a sufficiently high input resistance, with the only signal attenuation arising from parasitic input capacitance. Despite the significant differences in capacitor values, there is overall consistency in both gain and phase responses that the sensor can achieve.
\begin{table*}[h!]
    %\color{blue}
    \centering
    \begin{threeparttable}
        \caption{State-of-the-Art Fetal Monitoring Systems Comparison}
        \label{tab:state-of-the-art}
        
        \renewcommand{\arraystretch}{1.3} 
        \small % Use a slightly smaller font for the whole table
        % Use tabularx to make the table exactly \textwidth wide by wrapping text in X columns.
        % >{\centering\arraybackslash}X makes the wrapping columns centered.
        \begin{tabularx}{\textwidth}{l *{6}{>{\centering\arraybackslash}X}}
            \toprule
            \textbf{Comparison Criteria} & \textbf{\cite{SOTA_01}} & \textbf{\cite{SOTA_02}} & \textbf{\cite{Zhang_et_al}} & \textbf{\cite{SOTA_04}} & \textbf{\cite{SOTA_05}} & \textbf{This study} \\ 
            \midrule
            Type of Electrodes & Non-contact & Non-contact & Contact & Contact & Contact & \textbf{Non-contact} \\ 
            Min. Coupling Capacitance (pF) & N/A & N/A & N/A & N/A & N/A & \textbf{5} \\ 
            Reported Gain (V/V) & 1000 & N/A & 24 & 3.5\tnote{a} & 24 & \textbf{264} \\
            Bandwidth (Hz) & 0.5--40 & N/A & N/A & N/A--110 & 3--100 & \textbf{0.07--250} \\
            ADC Unit & nRF51422 & ADS1299 & ADS1299 & ADS1293 & ADS1299 & \textbf{ADS1299} \\ 
            ADC Resolution (bit) & 10 & 24 & 24 & 24 & 24 & \textbf{24} \\
            Sampling Rate (Hz) & 200 & 500 & 500 & 250 & 500 & \textbf{500} \\ 
            Theoretical Sensitivity ($\mu$V/LSB)\tnote{b} & 1.17 & N/A & 0.022 & 0.086\tnote{c} & 0.022 & \textbf{0.002} \\
            Processing Unit & nRF51422 & nRF52832 & STM32F103 & CY8C4247 & STM32WB55 & \textbf{nRF52832} \\ 
            Number of Subjects: & & & & & & \\ 
            \quad \textit{Pregnant women} & 1 & 10 & 3 & 0 & 3 & \textbf{10} \\
            \quad \textit{Others} & 1 & 1 & 0 & 0 & 0 & \textbf{4} \\ 
            \bottomrule
        \end{tabularx}
        
        \begin{tablenotes}
            \small 
            \item[a] Inferred from the ADS1293 parameters and the reported dynamic input range of the measurement circuit.
            \item[b] Theoretical sensitivity represents the smallest change the acquisition circuit can detect, corresponding to 1 LSB (least significant bit) of the ADC (analog-to-digital converter).  We assume all ADCs are considered to operate with their nominal internal reference voltage.
            \item[c] Estimated from the distinctive output coding scheme of ADS1293.
        \end{tablenotes}
    \end{threeparttable}
\end{table*}
In the next experiment, noise levels were measured using the method described in the previous section and the input referred noise spectrum is shown in Fig. \ref{fig: Noise analysis}. The higher the source capacitor, the lower noise level is observed. The measured in-band input referred noise at 10 Hz is approximately $10^{-6} V_{rms}$ for both the source capacitor of 30 pF and 100 pF while it is slightly higher at 5 pF.

{Table \ref{tab:state-of-the-art} compares the proposed system with five of the most recent studies on fetal ECG measurement in pregnant women. Relative to contact-based designs, the proposed system demonstrates competitive characteristics. Specifically, it achieves the highest bandwidth (0.07–250 Hz) among all fully reported studies, covering even a wider spectrum than contact-based works. Furthermore, its theoretical sensitivity (0.002 $\mu$V/LSB) markedly surpasses all other reported values (e.g., 0.086 $\mu$V/LSB in \cite{SOTA_04} or 0.022 $\mu$V/LSB in \cite{Zhang_et_al} and \cite{SOTA_05}). Within the non-contact group, the proposed system is distinguished by a higher level of design maturity, being carefully validated with both healthy individuals and ten pregnant women. Importantly, it explicitly evaluates the minimum coupling capacitance (5 pF) required for maintaining nominal performance, an aspect that has not been addressed in earlier works. This factor is crucial in real-world scenarios where clothing introduces extremely weak electrode–skin coupling, significantly reducing capacitance. By guaranteeing system performance down to this capacitance threshold, the proposed design addresses a key limitation of prior non-contact studies}.
\subsection{Physiological recording with NCE on healthy subjects}
\begin{figure*}[!ht]
    \centering
    \includegraphics[width=\textwidth]{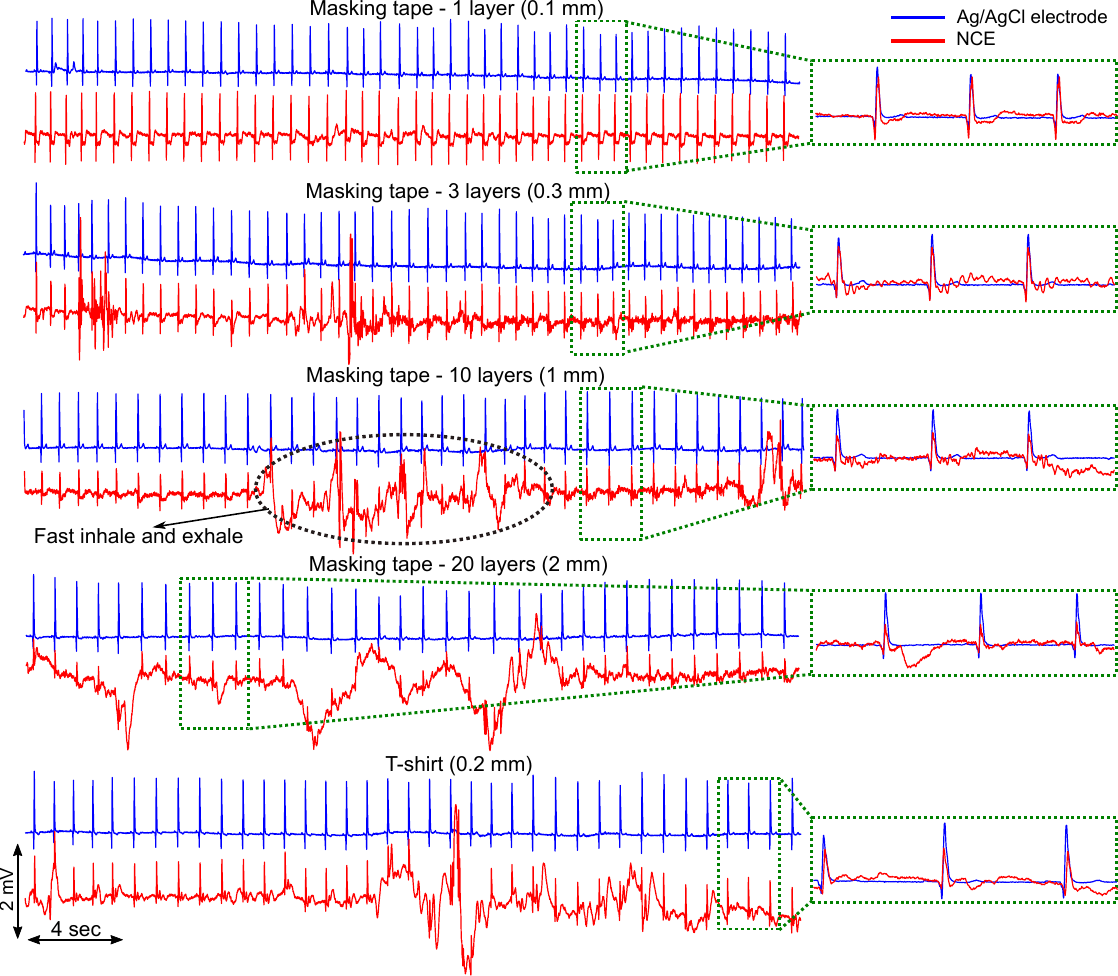}
    \caption{ECG representative on healthy subjects with different sensors’ insulator thickness.}
    \label{fig: ECG representative}
\end{figure*}

\begin{figure}[!ht]
    \centering
    \includegraphics[width=\linewidth]{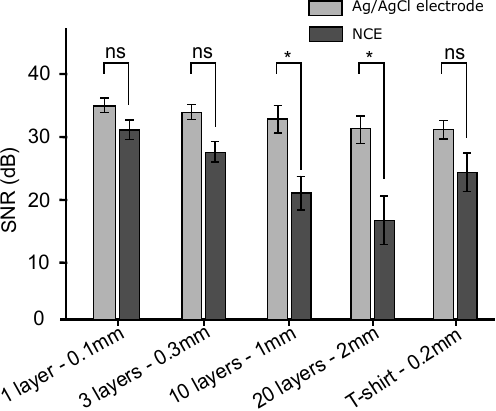}
    \caption{Signal-to-noise comparison among wet electrodes and NCE with different insulators’ thickness. ns indicates not significant.}
    \label{fig: SNR Different insulator thickness}
\end{figure}

Performing ECG measurement on health subjects with the NCE sensor, Fig. \ref{fig: ECG representative} depicts the ECG representative collected with different sensor's insulators and conventional electrode - Ag/AgCl. The ECG signal produced with the NCE sensor using 0.1 mm insulator of masking tape showed comparable in terms of signal quality with Ag/AgCl electrode. Applying thicker insulation layer, the NCE sensor is more susceptible to motion artifact and other interference. For instance, a volunteer subject performing rapid breathing exhibited ECG signals dominated by motion noise when using a NCE sensor with a 1 mm thick insulator, in contrast to the cleaner signal obtained with a wet electrode. Nevertheless, as shown in the right side with inset figures, the QRS complex is adequate to identify with ECG data collected by the NCE sensor. Fig. \ref{fig: SNR Different insulator thickness} illustrates a comparative analysis of the SNR between Ag/AgCl electrodes and NCE sensor under varying insulation conditions. Results demonstrate that NCEs with thinner insulation layers (0.1 mm and 0.3 mm) exhibited SNRs statistically comparable to those of Ag/AgCl electrodes, as did measurements taken through a standard T-shirt (0.2 mm). Conversely, a significant reduction in SNR was observed for NCEs with thicker insulation (1 mm and 2 mm), suggesting an inverse relationship between insulation thickness and signal quality in non-contact ECG measurements

\begin{figure}[!ht]
    \centering
    \includegraphics[width=\linewidth]{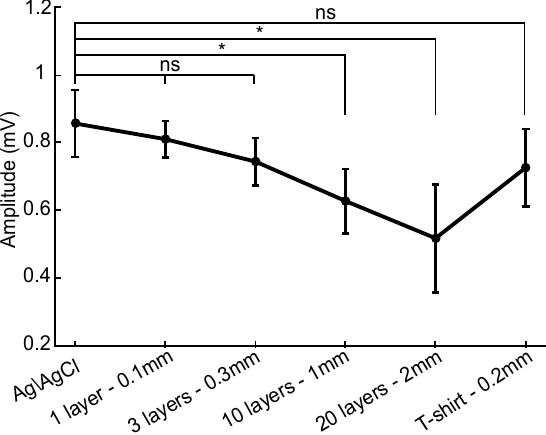}
    \caption{Amplitude comparison among wet electrodes and NCE with different insulators’ thickness. ns indicates not significant.}
    \label{fig: amplitude comparison}
\end{figure}

Further characterization of the R-peak amplitudes of the ECG signal, as recorded from healthy subjects and presented in Fig. \ref{fig: amplitude comparison}, was conducted. The NCE sensor with mask tape insulators of 0.1 mm and 0.3 mm thickness yielded signal strengths comparable to those obtained with the Ag/AgCl electrodes, demonstrating no statistically significant difference. However, increasing the insulation thickness to 1 mm and 2 mm resulted in a significant reduction in signal amplitude, measuring approximately 0.62 mV and 0.51 mV, respectively, in comparison to the 0.85 mV observed with the Ag/AgCl electrodes. Notably, no significant difference was observed between measurements acquired through a standard T-shirt and those obtained using wet electrodes. Given that the majority of common clothing exhibits thicknesses analogous to that of a T-shirt, the present study employed considerably thicker insulators (e.g., 1 mm and 2 mm) to deliberately stress the NCE sensor, thereby evaluating its performance under more challenging conditions.

\begin{figure}[!ht]
    \centering
    \includegraphics[width=\linewidth]{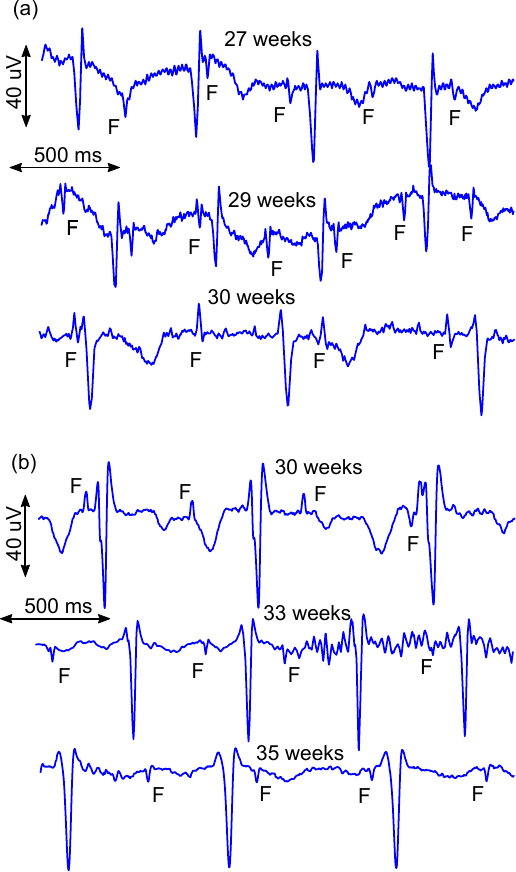}
    \caption{Fetal/maternal ECG representative from two subjects with different gestational ages in the pilot study.}
    \label{fig: Pregnant subjects with diff gestational ages}
\end{figure}

\subsection{Physiological recording with NCE on pregnant subjects}
Fig. \ref{fig: Pregnant subjects with diff gestational ages} illustrates the mixed fetal and maternal ECG collected from two subjects at various gestational ages using our system with non-contact electrodes (NCEs). The larger R peaks associate with maternal QRS complexes, while the smaller R peaks (also denoted as “F” in the figure) belong to the fetus. The NCE was capable of picking up fetal QRS complexes throughout different weeks of pregnancy, demonstrating its potential in fetal ECG monitoring.
\begin{table*}[!htbp]
\centering
\caption{Comparison of fmECG signal quality and fQRS detection performance by using the novel NCE and the Ag/AgCl electrodes on pregnant subjects} \label{tab:evaluation}
\begin{tabular}{ccccccccccc}
\toprule
Subject &  \multicolumn{2}{c}{SNR (dB)} &\multicolumn{2}{c}{Se(\%)} & \multicolumn{2}{c}{PPV(\%)} & \multicolumn{2}{c}{ACC(\%)} & \multicolumn{2}{c}{F1(\%)}\\
\midrule
{}   & Ag/AgCl   & NCE & Ag/AgCl   & NCE    & Ag/AgCl   & NCE & Ag/AgCl   & NCE & Ag/AgCl   & NCE \\
1 &                29.41 &            26.95 &            96.58 &        90.46 &             95.01 &         94.37 &             95.86 &         93.20 &            95.79 &        92.37 \\
2 &                27.52 &            25.10 &            96.65 &        93.82 &             96.71 &         92.55 &             95.87 &         91.21 &            96.68 &        93.18 \\
3 &                27.78 &             \textbf{14.05} &            96.68 &        \textbf{54.31} &             95.97 &         \textbf{53.20} &             95.32 &         \textbf{53.38} &            96.32 &        \textbf{53.74} \\
4 &                27.65 &            25.22 &            96.11 &        94.06 &             95.70 &         89.97 &             96.31 &         89.71 &            95.90 &        91.97 \\
5 &                28.14 &            27.24 &            95.73 &        94.31 &             96.17 &         93.01 &             96.91 &         89.38 &            95.95 &        93.66 \\
6 &                25.64 &            26.34 &            95.61 &        93.23 &             96.62 &         90.39 &             95.25 &         94.39 &            96.11 &        91.79 \\
7 &                28.63 &            26.08 &            95.99 &        91.80 &             96.85 &         92.02 &             96.02 &         89.40 &            96.42 &        91.91 \\
8 &                28.44 &            \textbf{17.10} &            96.47 &        \textbf{61.70} &             96.02 &         \textbf{64.34} &             95.47 &         \textbf{62.30} &            96.24 &        \textbf{63.00} \\
9 &                25.14 &            24.87 &            95.70 &        91.71 &             95.18 &         93.98 &             95.75 &         93.73 &            95.44 &        92.83 \\
10 &                28.59 &            26.15 &            96.05 &        94.02 &             96.17 &         93.97 &             95.49 &         91.01 &            96.11 &        93.99 \\
\midrule
Average &                27.69 &            23.91 &            96.16 &        85.94 &             96.04 &         85.78 &             95.83 &         84.71 &            96.10 &        85.84 \\
\bottomrule
\end{tabular}
\end{table*}
For comparison, we further characterized the signal quality using SNR and fQRS detection accuracy between NCEs and conventional Ag/AgCl electrodes in Table \ref{tab:evaluation}. Overall, the fetal maternal ECG (fmECG) data recorded by Ag/AgCl electrodes were less noisy, achieving a peak SNR of 29.41 dB. The NCEs yielded a lower maximum SNR of 27.24 dB and exhibited greater performance variability across subjects, especially in subjects 3 and 8 with significantly poorer outcomes. While the average F1 score for fQRS detection using Ag/AgCl electrodes was a consistent 96.10\%, the scores in subjects 3 (53.74\%) and 8 (63\%) with NCEs were abysmal compared to the rest (92.7\% on average), highlighting the specific challenges of NCE-based measurements in certain conditions.
{The notably poor signal quality and detection accuracy for subjects 3 and 8 can be attributed to several physiological and technical factors known to affect abdominal fECG recordings, especially with NCEs. First, both subjects were in later gestational weeks (28-32 weeks) with the strong development of the vernix caseosa. This low-conductivity layer forms on the fetus’s skin which significantly attenuates the electrical signal \cite{Sameni_2020}. NCEs are sensitive to changes in electrical impedance and thus could be affected disproportionately by vernix caseosa. Second, the fetal position can be in an unfavorable orientation with fetal cardiac electrical axes (pointing toward the mother’s back) facing away from the electrode locations (on the front), reducing signal amplitude at the skin’s surface. Finally, maternal movement and respiration can introduce motion artifacts, which increase signal interference and further weaken the capacitively coupled signal \cite{behar2016extractionclinicalinformationnoninvasive, Sameni_2020}. The NCE's sensitivity to distance and coupling makes it more susceptible to these issues compared to adhesive Ag/AgCl electrodes}.

{It is also critical to acknowledge the limitations of this pilot feasibility study. The small sample size (n=10) precludes definitive clinical claims and robust statistical analysis of performance across different gestational ages or maternal. Furthermore, while subjects were seated comfortably, this study did not include a quantitative analysis of motion artifacts, which is essential for assessing real-world performance in an ambulatory home-monitoring scenario. The electrode placement strategy was also not optimized using anatomical landmarks or ultrasound, which could have improved signal acquisition}.

Throughout the pilot study, the standard electrode caused skin irritation and left some residual conductive gel, which is hard to remove. In contrast, the use of NCE brings comfort to users as it does not come into direct contact with the user’s skin. This would bring benefits for the long-term usage of the device.

\section{Conclusion}
In this study, a fetal ECG monitoring system with the use of novel NCE is developed. Various experiments were performed to validate the operation of all components as well as evaluate its performance against the Ag/AgCl electrodes. The results revealed that the quality of the ECG data collected by the NCE was found to be comparable with the established standard of Ag/AgCl electrodes in volunteer subjects. The fECG monitoring system was tested on 10 pregnant subjects at UCI Medical Center, demonstrating its performance and usability.

{In the future, we plan to incorporate different sensors to locate the fetus location and orientation so that it helps with the NCE placement, thereby improve the signal quality.} {We will also investigate advanced signal processing algorithms for motion artifact removal}. Moreover, the data collection of more participants will be carried out to verify its performance and feasibility for out-of-clinics fECG monitoring in daily life.

\newpage
\printbibliography % Print References
\newpage

\section{Biography Section}

\vfill

\end{document}